\documentclass[prc,aps,twocolumn,showpacs]{revtex4}
\usepackage{graphicx}
\def\pcomma{$^,$}
\def\pucla{$^1$}
\def\pmainz{$^2$}
\def\pglsg{$^3$}
\def\pkent{$^4$}
\def\pbonn{$^5$}
\def\pgatch{$^6$}
\def\pgiess{$^7$}
\def\pbasel{$^8$}
\def\ppavia{$^9$}
\def\pedinb{$^{10}$}
\def\pgwu{$^{11}$}
\def\plpi{$^{12}$}
\def\psackv{$^{13}$}
\def\pinr{$^{14}$}
\def\pzagreb{$^{15}$}
\def\pcua{$^{16}$}

\begin{document}

\title{Measurement of the Slope Parameter $\alpha$
 for the $\eta\to 3\pi^0$ decay with the Crystal Ball detector
 at the Mainz Microtron (MAMI-C)}

\author{
S.~Prakhov\pucla\footnote[1]{Electronic address: prakhov@ucla.edu}, 
B.~M.~K.~Nefkens\pucla,
P.~Aguar-Bartolom\'e\pmainz,
L.~K.~Akasoy\pmainz,
J.~R.~M.~Annand\pglsg,
H.~J.~Arends\pmainz,
K.~Bantawa\pkent,
R.~Beck\pmainz\pcomma\pbonn,
V.~Bekrenev\pgatch,
H.~Bergh\"auser\pgiess,
B.~Boillat\pbasel,
A.~Braghieri\ppavia,
D.~Branford\pedinb,
W.~J.~Briscoe\pgwu,
J.~Brudvik\pucla,
S.~Cherepnya\plpi,
R.~F.~B.~Codling\pglsg,
E.~J.~Downie\pmainz\pcomma\pglsg,
P.~Drexler\pgiess,
L.~V.~Fil'kov\plpi,
D.~I.~Glazier\pedinb,
R.~Gregor\pgiess,
E.~Heid\pmainz\pcomma\pgwu,
D.~Hornidge\psackv,
O.~Jahn\pmainz,
T.~C.~Jude\pedinb,
V.~L.~Kashevarov\plpi,
J.~D.~Kellie\pglsg,
R.~Kondratiev\pinr,
M.~Korolija\pzagreb,
M.~Kotulla\pgiess,
A.~Koulbardis\pgatch,
D.~Krambrich\pmainz,
S.~Kruglov\pgatch,
B.~Krusche\pbasel,
M.~Lang\pmainz\pcomma\pbonn,
V.~Lisin\pinr,
K.~Livingston\pglsg,
I.~J.~D.~MacGregor\pglsg,
Y.~Maghrbi\pbasel,
D.~M.~Manley\pkent,
M.~Martinez\pmainz,
J.~C.~McGeorge\pglsg,
E.~F.~McNicoll\pglsg,
D.~Mekterovic\pzagreb,
V.~Metag\pgiess,
S.~Micanovic\pzagreb,
A.~Nikolaev\pbonn,
R.~Novotny\pgiess,
M.~Ostrick\pmainz,
P.~B.~Otte\pmainz,
P.~Pedroni\ppavia,
F.~Pheron\pbasel,
A.~Polonski\pinr,
J.~Robinson\pglsg,
G.~Rosner\pglsg,
M.~Rost\pmainz,
T.~Rostomyan\ppavia\footnote[2]{Present address:
 Institut f\"ur Physik, University of Basel, Switzerland},
S.~Schumann\pmainz\pcomma\pbonn,
M.~H.~Sikora\pedinb,
D.~I.~Sober\pcua,
A.~Starostin\pucla,
I.~M.~Suarez\pucla,
I.~Supek\pzagreb,
C.~M.~Tarbert\pedinb,
M.~Thiel\pgiess,
A.~Thomas\pmainz,
M.~Unverzagt\pmainz\pcomma\pbonn,
D.~P.~Watts\pedinb,
I.~Zamboni\pzagreb,
and F.~Zehr\pbasel
 \\
\vspace*{0.1in}
(Crystal Ball Collaboration at MAMI and A2 Collaboration)
\vspace*{0.1in}
}

\affiliation{
\pucla University of California Los Angeles, Los Angeles,
 California 90095-1547, USA}
\affiliation{
\pmainz Institute f\"ur Kernphysik, University of Mainz, D-55099 Mainz, Germany}
\affiliation{
\pglsg Department of Physics and Astronomy, University of Glasgow, Glasgow G12 8QQ, United Kingdom}
\affiliation{
\pkent Kent State University, Kent, Ohio 44242-0001, USA}
\affiliation{
\pbonn Helmholtz-Institut f\"ur Strahlen- und Kernphysik, University of Bonn, D-53115 Bonn, Germany}
\affiliation{
\pgatch Petersburg Nuclear Physics Institute, 188350 Gatchina, Russia}
\affiliation{
\pgiess II Physikalisches Institut, University of Giessen, D-35392 Giessen, Germany}
\affiliation{
\pbasel Institut f\"ur Physik, University of Basel, CH-4056 Basel, Switzerland}
\affiliation{
\ppavia INFN Sesione di Pavia, I-27100 Pavia, Italy}
\affiliation{
\pedinb School of Physics, University of Edinburgh, Edinburgh EH9 3JZ, United Kingdom}
\affiliation{
\pgwu The George Washington University, Washington, DC 20052-0001, USA}
\affiliation{
\plpi Lebedev Physical Institute, 119991 Moscow, Russia}
\affiliation{
\psackv Mount Allison University, Sackville, New Brunswick E4L3B5, Canada}
\affiliation{
\pinr Institute for Nuclear Research, 125047 Moscow, Russia}
\affiliation{
\pzagreb Rudjer Boskovic Institute, HR-10000 Zagreb, Croatia}
\affiliation{
\pcua The Catholic University of America, Washington, DC 20064, USA}

\date{\today}

\begin{abstract}
 The dynamics of the $\eta\to 3\pi^0$ decay have been
 studied with the Crystal Ball multiphoton spectrometer
 and the TAPS calorimeter. Bremsstrahlung photons produced
 by the 1.5-GeV electron beam of the Mainz microtron MAMI-C
 and tagged by the Glasgow photon spectrometer were used
 for $\eta$-meson production.
 The analysis of $3 \times 10^6$
 $\gamma p \to \eta p \to 3\pi^0 p \to 6\gamma p$ events
 yields the value $\alpha=-0.032\pm0.003$ for the $\eta\to 3\pi^0$
 slope parameter, which agrees with the majority of recent
 experimental results and has the smallest uncertainty.
 The $\pi^0\pi^0$ invariant-mass spectrum was investigated
 for the occurrence of a cusplike structure
 in the vicinity of the $\pi^+\pi^-$ threshold.
 The observed effect is small and does not affect
 our measured value for the slope parameter.
\end{abstract}


\maketitle

\section{Introduction}

 The experimental study of the simple and pure
 strong-interaction reaction
\begin{equation}
 \pi^0 \pi^0 \to \pi^0 \pi^0 \label{eqn:2pi0int}
\end{equation}
 is a real challenge as neither a $\pi^0$ target nor a $\pi^0$ beam
 is available. The properties of reaction~(\ref{eqn:2pi0int})
 can be extracted indirectly from complicated
 processes, for example, from $K^+ \to \pi^0 \pi^0 e^+ \nu_e$ ($K^+_{e4}$),
 which is the weak decay of the $K^+$ followed by strong final-state
 interactions between the two $\pi^0$s.
 Major disadvantages of studying reaction~(\ref{eqn:2pi0int})
 in $K^+_{e4}$ are the small branching
 ratio ($2.2\times 10^{-5}$) and the complications from
 the four complex form factors for the $K^+_{e4}$
 decay amplitude needed to describe the four-particle
 final state.
\begin{figure*}
\includegraphics[width=13.cm,height=6.cm,bbllx=0.5cm,bblly=0.5cm,bburx=19.cm,bbury=10.5cm]{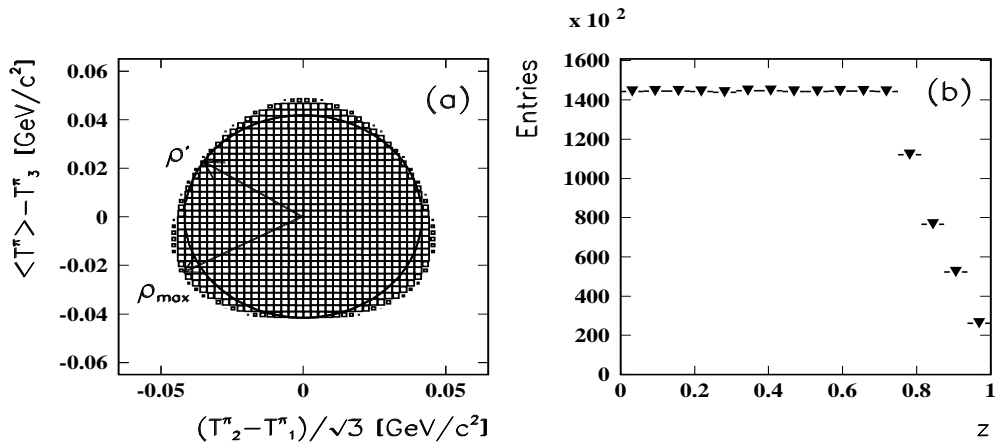}
\caption{
 Distributions for the phase-space decay of $\eta\to 3\pi^0$
 (i.e., when $\alpha=0$) obtained from a Monte Carlo simulation:
 (a) Dalitz plot, where variable $T_i^{\pi}$ is the kinetic energy of each
 of the three pions and $\langle T^{\pi} \rangle$
 is the mean kinetic energy
 of the three pions (with all energies being calculated in
 the $\eta$ rest frame);
 (b) variable $z=\rho^2/\rho^2_{\mathrm{max}}$, reflecting the
    density of the Dalitz plot.
}
 \label{fig:dk_eta3pi0_zeta} 
\end{figure*}
 Another process that can be used for the indirect study
 of reaction~(\ref{eqn:2pi0int}) is the decay
\begin{equation}
 \eta\to 3\pi^0~, \label{eqn:deta3pi0}
\end{equation}
 where the $\pi^0 \pi^0$ final-state interaction can be seen
 in a difference of the $\eta\to 3\pi^0$ decay amplitude
 from phase space.
 The experimental study of this decay has several
 major advantages: 
 the relatively large branching ratio for $\eta\to 3\pi^0$ (32.5\%),
 a high yield of $\eta$ mesons in many production
 reactions, and very small background from other $3\pi^0$ contributions,
 especially in $\eta$ production close to the threshold.

 Due to the low energies of the decay $\pi^0$s,
 $\pi^0\pi^0$ rescattering in $\eta\to 3\pi^0$
 is expected to be dominated by S and P waves.
 This leads to the parametrization of
 the $\eta\to 3\pi^0$ decay amplitude as
 $A(\eta\to 3\pi^0) \sim 1+ \alpha z$~\cite{PDG},
 where $\alpha$ is the quadratic slope parameter that describes the
 difference from phase space.
 A convenient definition of the kinematic variable $z$ is 
\begin{equation}
 z = 6\sum_{i=1}^3 (E_i - m_{\eta}/3)^2/(m_\eta-3m_{\pi^0})^2
  = \rho^2/\rho^2_{\mathrm{max}}~,  \label{eqn:zvar}
\end{equation}
 where $E_i$ is the energy of the $i$th pion 
 in the $\eta$ rest frame, and $\rho$ is the distance from the center
 of the $\eta\to 3\pi^0$ Dalitz plot.
 The variable $z$ varies from 0,
 when all three $\pi^0$s have the same energy of $m_{\eta}/3$,
 to 1, when one $\pi^0$ is at rest.
 A geometrical interpretation of Eq.~(\ref{eqn:zvar}) gives
 $z=0$ when $\rho=0$ and $z=1$ when $\rho=\rho_{\mathrm{max}}$.
 The density of the $\eta\to 3\pi^0$ Dalitz plot is
 described by $|A(\eta\to 3\pi^0)|^2 \sim 1+ 2\alpha z$.  
 The phase-space decay of $\eta\to 3\pi^0$  (i.e., when $\alpha=0$)
 gives a uniform density of the Dalitz plot,
 as shown in Fig.~\ref{fig:dk_eta3pi0_zeta}(a).
 The corresponding distribution of the variable $z$ is shown
 in Fig.~\ref{fig:dk_eta3pi0_zeta}(b);
 it is uniform for $z$ from 0 to $\approx$0.75~.
 Experimentally, the slope parameter $\alpha$ is usually determined from
 the deviation of the measured $z$ distribution from the corresponding
 distribution obtained by a Monte Carlo simulation in which
 the $\eta\to 3\pi^0$ decay amplitude is independent of $z$.
 
 The $\eta\to 3\pi^0$ decay, which violates G parity,
 occurs mostly because of the $u$-$d$ quark mass difference.
 The precision measurement of the $\eta\to 3\pi^0$ decay width,
 $\Gamma(\eta \to 3\pi^0) \sim (m_d-m_u)^2(1+2\alpha z)$,
 and the parameter $\alpha$
 are important tests of
 Chiral Perturbation Theory ($\chi$PTh)~\cite{gasser,kambor}.  
 In the $\chi$PTh momentum expansion in orders of the $\chi$PTh
 parameter $\sf p$,
 the leading ${\cal O}({\sf p}^2)$ term of the decay amplitude
 explicitly depends on $m_d-m_u$.
 However, including this term and
 the second-order counter terms, ${\cal  O}({\sf p}^4)$,
 is not sufficient~\cite{gasser} to yield a decay width
 that is close to the measured value of 423~eV~\cite{PDG}.
 The use of dispersion relations~\cite{kambor,anisovich},
 which include pion rescattering to all orders,
 partially improves the agreement with the experimental value.
 For the parameter $\alpha$, the dispersion-relation calculations
 of Ref.~\cite{kambor} give a negative value in the range
 $-0.007$ to $-0.0014$, depending on the assumptions made.
 The results of these calculations are outside the value of
 $\alpha=-0.031\pm0.004$ adopted by the Particle Data Group
 (PDG)~\cite{PDG}.
 This value for $\alpha$ is based on the analysis of $0.9\times10^6$
 $\eta\to 3\pi^0$ decays measured by the Crystal Ball
 at the AGS~\cite{eta_slope_bnl}.
 A complete two-loop calculation in standard $\chi$PTh~\cite{Bijnens}
 results in $\alpha=0.013\pm0.032$, the sign of which is even
 opposite to the experimental values.
 The evaluation of the electromagnetic corrections in
 the $\eta\to 3\pi^0$ decay~\cite{Ditsche} shows that
 they are too small to explain the difference
 between the $\chi$PTh calculations for $\alpha$
 and the experimental results.
 Only the use of a chiral unitary approach based on
 the Bethe-Salpeter equation~\cite{Borasoy} yields
$\alpha=-0.031\pm0.003$, which is in very good agreement
 with the PDG value.
 Several new experiments, which aim to remeasure $\alpha$ with better
 statistics, are still under way.
 So far, the latest preliminary result from
 the KLOE Collaboration~\cite{KLOE2},
 $\alpha=-0.027\pm0.004_{\mathrm{stat}}~^{+0.004}_{-0.006\mathrm{syst}}$,
 is based on poorer statistics 
 ($0.65\times10^6$ $\eta\to 3\pi^0$ decays) and is in 
 agreement with the PDG value within the errors.

 The experimental study of the $\eta \to 3\pi^0$ decay
 has recently become of special interest because of new results
 of the NA48/2 Collaboration~\cite{NA48} that were obtained
 from the analysis of $K^+ \to \pi^+\pi^0\pi^0$ decays,
 where a significant cusp effect was observed
 in the $\pi^0\pi^0$ invariant-mass
 spectrum close to the $\pi^+\pi^-$ threshold.    
 The cusp occurs because
 the $K^+ \to \pi^+\pi^+\pi^-$ decay contributes via
 the $\pi^+\pi^- \to \pi^0\pi^0$ charge exchange reaction to
 the $K^+ \to \pi^+\pi^0\pi^0$ decay amplitude.
 Cusps in $\pi\pi$ scattering were described
 in the $\chi$PTh framework  
 by Meissner {\it et al.}~\cite{Meissner}. 
 The cusp characteristics were used for the experimental
 determination of the $\pi\pi$ scattering length
 combination $a_0 - a_2$, the $\chi$PTh prediction for which
 is $0.265\pm 0.004$~\cite{Colangelo}.
 The method for the determination of $a_0 - a_2$
 from the analysis of the $\pi^0\pi^0$ invariant-mass spectrum
 from the $K^+ \to \pi^+\pi^0\pi^0$ decays
 has been presented by Cabibbo~\cite{Cabibbo}.
 A cusp effect in the $\eta \to 3\pi^0$ decay,
 arising because of the $\eta \to \pi^+\pi^-\pi^0$ decay contribution,
 is expected to be less significant~\cite{Belina}.
 This makes it less attractive
 for the experimental extraction of the $\pi\pi$ scattering lengths,
 but neglecting the cusp effect in the analysis of the $z$
 distribution could result in the wrong experimental value for
 $\alpha$. In a situation like this, a new, high-statistics measurement of
 the $\eta \to 3\pi^0$ decays with good resolution in
 the $\pi^0\pi^0$ invariant mass and in the variable $z$
 is desirable.
 
 In this paper, we report on a new precision measurement
 of the slope parameter $\alpha$ for the $\eta \to 3\pi^0$ decay
 that was made by the Crystal Ball Collaboration at MAMI-C.
 These data are also used to look for
 a cusp structure in the $\pi^0\pi^0$ invariant-mass spectrum
 and for estimating how the cusp can affect the result for $\alpha$.
 There is also an independent analysis of the $\eta \to 3\pi^0$
 data taken by the Crystal Ball with a lower beam energy
 of MAMI-B~\cite{MarcTh,MarcAr}.
 The result for $\alpha$ reported there is in good agreement with the
 present work.  
   
\section{Experimental setup}
 
 The study of the $\eta \to 3\pi^0$ decay was done by
 measuring the process $\gamma p \to \eta p \to 3\pi^0 p \to 6\gamma p$
 with the Crystal Ball (CB) multiphoton spectrometer~\cite{etalam}
 used as the central detector
 and the TAPS calorimeter~\cite{TAPS,TAPS2} as the forward detector.
 The experimental setup was installed
 in the bremsstrahlung photon beam of the Mainz Microtron (MAMI)~\cite{MAMI,MAMIC},
 with the photon energies determined by
 the Glasgow tagging spectrometer~\cite{Anthony,Hall,Tagger2}.

 The Crystal Ball spectrometer was originally built by SLAC
 for studies of $e^+e^-$ collisions in the $J/\psi$ region~\cite{Oreglia}.
 The recent history of the CB starts at the AGS where
 the first high-statistics
 measurement of the $\eta \to 3\pi^0$ slope parameter
 was carried out with the $\eta$-production
 reaction $\pi^- p \to \eta n$~\cite{eta_slope_bnl}.
 After a stint at the AGS, the CB was moved in 2002
 to the Mainz Microtron for a large variety of experiments,
 including studies of $\eta$ photoproduction and rare
 and special decay modes of $\eta$.

 The Crystal Ball spectrometer is a sphere
 consisting of 672 optically insulated NaI(Tl) crystals,
 shaped as truncated triangular pyramids,
 all pointing towards the center of the CB.
 The crystals are
 arranged in two hemispheres that cover 93\% of $4\pi$ steradians.
 The CB has a spherical cavity in the center with radius of 25~cm;
 it is designed to hold a target and inner detectors.
 There are also two tunnels shaped close to a $40^\circ$ cone;
 they serve for entrance and exit of the beam.
 Each NaI(Tl) crystal is 41~cm long, which
 corresponds to 15.7 radiation lengths.   
 As the decay time of NaI(Tl) is about 250~ns, high count rates
 cause pulse pileup and worsen the energy resolution.
 For the runs with the normal count rate,
 the energy resolution for electromagnetic showers in the CB 
 can be described as $\Delta E/E = 0.020/(E[\mathrm{GeV}])^{0.36}$. 
 Shower directions are measured with a resolution in $\theta$,
 the polar angle with respect to the beam axis, 
 of $\sigma_\theta = 2^\circ \textrm{--} 3^\circ$,
 under the assumption that the photons are
 produced in the center of the CB. The resolution in
 the azimuthal angle $\phi$ is $\sigma_\theta/\sin\theta$.
 More details on the CB spectrometer and the physics recently studied with it
 at the AGS can be found in
 Refs.~\cite{n2pi0,Sadler,l2pi0,etapi02g}.
 For experiments at MAMI, the CB has been equipped with
 new electronics, providing
 an individual TDC (time-to-digital converter) for each crystal.
 This gives better suppression
 of pileups in the analysis of the MAMI data in comparison
 with earlier CB experiments, where one TDC reads out nine crystals.
  
 To cover the downstream beam tunnel of the CB,
 the TAPS calorimeter~\cite{TAPS,TAPS2}
 was installed 1.5~m downstream of the CB center.
 TAPS is designed as a versatile end-plane hodoscope that can be
 arranged in different formations to optimize the detection of 
 the forward-going final-state particles.
 In this experiment, TAPS was arranged in a plane 
 consisting of 384 individual BaF$_2$ counters
 that are hexagonally shaped with an inner diameter of 5.9~cm and a length
 of 25~cm, which corresponds to 12 radiation lengths.
 The beam hole of the TAPS calorimeter has a shape of one BaF$_2$ counter
 removed from the hodoscope center. 
 The energy resolution for electromagnetic showers in the TAPS calorimeter
 can be described as $\Delta E/E = 0.018 + 0.008/(E[\mathrm{GeV}])^{0.5}$.
 Because of the long distance from the CB, the resolution of TAPS
 in the polar angle $\theta$ was better than $1^\circ$.
 The resolution of TAPS in the
 azimuthal angle $\phi$ is better than $1/R$ radian, where $R$ is 
 the distance in centimeters from the TAPS center to the point on the TAPS surface
 that corresponds to the $\theta$ angle. This means that
 the azimuthal-angle resolution of TAPS becomes better than $1^\circ$
 when $R>57$~cm.
    
 The upgraded Mainz Microtron, MAMI-C, is a harmonic double-sided electron
 accelerator with a maximum beam energy of 1508~MeV~\cite{MAMIC}.
 The bremsstrahlung photons, produced by the electrons
 in a 10-$\mu$m Cu radiator and collimated by a 4-mm-diameter
 Pb collimator,
 were incident on a 5-cm-long liquid hydrogen (LH$_2$)
 target located in the center of the CB.
 The incident photons were tagged up to the energy of 1402~MeV
 using the post-bremsstrahlung electrons detected by the Glasgow
 tagger~\cite{Anthony,Hall,Tagger2}.
 The tagger consists of a momentum-dispersing magnetic spectrometer
 that focuses the electrons onto the focal plane detector
 of 353 half-overlapping plastic scintillators.
 The energy resolution of the tagged photons is mostly defined
 by the overlap region of two adjacent scintillation counters
 (a tagger channel) and the electron beam energy.
 For the electron beam of 1508~MeV,
 the tagger channel has a width about 2~MeV at 1402~MeV
 (the maximum of the tagging range) and about 4~MeV at 707~MeV
 (the $\eta$-production threshold).
 In the analysis, every photon is characterized
 by the time difference (tagging time)
 between the signal in the corresponding tagger channel
 and the experimental trigger.

 The LH$_2$ target is surrounded by a Particle IDentification
 (PID) detector~\cite{PID} that is
 a 50-cm-long, 12-cm-diameter cylinder built
 of 24 4-mm-thick plastic scintillators,
 which identify charged particles.
 The PID detector was not used in the present analysis.

 The experimental (or DAQ) trigger had two main requirements.
 First, the sum of the pulse amplitudes from the CB crystals had to exceed
 the hardware threshold that corresponded to an energy deposit of 320~MeV.
 Second, the number of ``hardware'' clusters in the CB
 had to be larger than one. In the trigger, a ``hardware'' cluster is 
 a block of 16 adjacent crystals in which at least one crystal
 has an energy deposit larger than 30~MeV.

 A technical paper that will describe more details on features,
 calibrations, and resolutions of our experimental setup
 is in preparation.

\section{Data analysis}

 The $\eta \to 3\pi^0$ decays were measured by the process
\begin{equation}
  \gamma p \to \eta p \to 3\pi^0 p \to 6\gamma p  \label{eqn:eta3pi0}
\end{equation}
 that was extracted by the analysis of events having
 six and seven ``software'' clusters reconstructed in the CB and TAPS together.
 The six-cluster sample was used to search for events of
 reaction~(\ref{eqn:eta3pi0}) in which only six photons were detected,
 and the proton went undetected. 
 The seven-cluster sample was used to search for events
 in which all six photons and the proton were detected in the CB and TAPS.
 The cluster algorithm in software was optimized
 for finding a group of adjacent crystals in which the energy was 
 deposited by a single-photon electromagnetic shower.
 This algorithm also works well for a proton cluster.
 The software threshold for the cluster energy was chosen to be 12~MeV.
 This value optimizes the number of good events reconstructed
 for the major processes:
 $\gamma p \to \eta p \to 3\pi^0 p$, $\gamma p \to \eta p \to \gamma\gamma p$,
 $\gamma p \to \pi^0 p$, and $\gamma p \to \pi^0 \pi^0 p$ . 
 The hardware read-out thresholds for individual crystals
 were about 1.1~MeV for the CB and 3.3~MeV for TAPS.
 To match the experimental data, corresponding software thresholds
 were introduced in the analysis of the Monte Carlo (MC) simulation.

 The kinematic-fitting technique~\cite{Blobel}
 was used to test various reaction hypotheses needed in
 our analysis.
 In our kinematic fit, the incident photon is parametrized
 by three measured variables: energy
 and event-vertex coordinates $X$ and $Y$ in the target.
 The direction of the incident photon is assumed to be parallel to the $Z$ axis.
 The initial values for the mean of $X$ and $Y$ are set
 to zero. The uncertainties $\sigma_X$ and $\sigma_Y$ are
 taken as the rms (root mean square) of the $X$ and $Y$
 distribution on the target. Their magnitudes, which
 typically are of a few millimeters,
 are determined by the collimator diameter and
 the 2.5-m distance between the collimator and the target. 
 The uncertainty in the energy of the incident photon
 is taken as one-third of the energy width of the corresponding
 tagger channel; this is slightly larger than the rms
 of a uniform distribution. To reproduce the same
 conditions in the analysis of the MC simulation, the simulated
 photon energy is substituted for the energy of the corresponding
 tagger channel.
 The $Z$ coordinate of the vertex is a free variable
 in the kinematic fit.
 A photon cluster in the fit is parametrized by four measured variables.
 For the CB, they are the cluster energy,
 the angles $\theta$ and $\phi$ calculated with respect to the CB center,
 and the effective depth of the shower in the crystals.
 For TAPS instead of the $\theta$ angle, the measured variable
 is the distance $R$ (in centimeters) from the cluster center
 to the $Z$ axis (the same as the TAPS center).  
 In the fitting procedure, the $\theta$ angle of a photon
 is calculated from the variable $R$ of the cluster,
 the distance from the vertex to the TAPS surface, and
 the shower effective depth in TAPS.
 The effective shower depth in the CB and TAPS
 is defined as the distance from
 the crystal front surface to the point where the deposit
 of the particle energy equals half of the cluster energy.
 The energy dependence of the effective depth and its uncertainty
 for the photon and proton clusters were determined
 from the MC simulation.

 The resolution of the CB in the cluster angle $\theta$ as a function
 of the cluster energy was determined for photons and protons
 from the difference between initial and reconstructed angles in the
 MC simulation. 
 The resolution of the CB in the cluster angle $\phi$ can be obtained
 from the $\theta$ resolution by dividing
 it by $\sin\theta$ (i.e., $\sigma_{\phi}=\sigma_{\theta}/\sin\theta$).
 To determine the resolution of TAPS in the distance $R$
 as a function of the cluster energy, the initial and reconstructed values
 $R$ from the MC simulation were compared.
 The resolution of TAPS in the angle $\phi$
 can be obtained from the $R$ resolution by dividing
 it by $R$ itself (i.e., $\sigma_{\phi}=\sigma_{R}/R$).
 The resolutions so determined are used in the kinematic fit as
 the uncertainties in the cluster parameters
 $\theta$, $R$, and $\phi$.

 Besides the individual gain coefficients for each crystal,
 which are used to calculate the deposited energy from
 the ADC (analog-to-digital converter) channels,
 we introduced an energy-dependent function that provides
 a correction for the cluster
 energy to get the photon energy.
 The correction is energy dependent bacause
 less energetic final-state photons
 have a larger fraction of the energy deposit that is not
 collected from the crystals due to their read-out thresholds.
 The magnitude of the correction
 was determined from the MC simulation by comparing
 the energy of the simulated photons and their
 reconstructed clusters. The advantage of using this function
 is that it improves the invariant-mass resolution
 and removes the dependence of the mean invariant-mass value on
 the energy of the incident photon. 
 Without using this function,
 the peaks from $\pi^0$ and $\eta$ in the invariant-mass spectra
 move to higher masses when increasing the incident-photon energy;
 this is a consequence of more energetic final-state photons.
\begin{figure*}
\includegraphics[width=14.5cm,height=6.cm,bbllx=1.cm,bblly=0.5cm,bburx=19.cm,bbury=8.5cm]{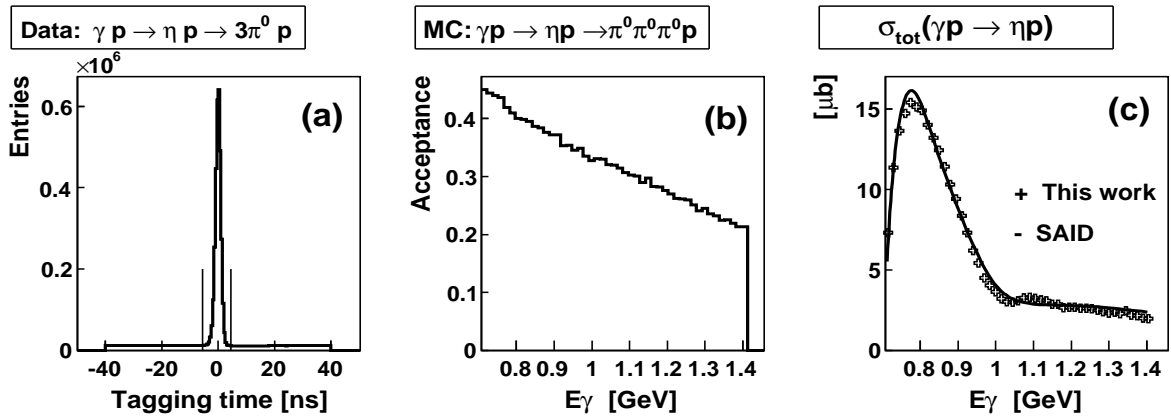}
\caption{
 (a) Tagging-time distribution for
    the experimental $\eta\to 3\pi^0$ events,
    where the two vertical lines define the prompt region.
 (b) Acceptance for the $\eta\to 3\pi^0$ events as a function
    of the incident-photon energy.
 (c) Measured $\gamma p \to \eta p$ excitation function
    compared to the current PWA-fit solution from
    the SAID~\protect\cite{SAID} data base.
}
 \label{fig:beameta_mamic} 
\end{figure*}

 The determination of the energy-resolution function $\Delta E/E$
 for the CB and TAPS was divided into two steps.
 First, we fixed a so-called inherent resolution of the
 calorimeters, which is determined only by the propagation
 of electromagnetic showers in the {\sc GEANT} simulation
 and by the cluster-reconstruction algorithm.
 The inherent energy resolution is better than the experimental one, which
 also includes additional smearing from the light collection, the conversion
 to electronic signals in the PMTs (photomultiplier tubes),
 digitization in the ADCs, and pileups.
 To match the experimental resolution, which is correlated
 with the beam intensity and depends on the quality of the gain calibrations,
 the MC output for the energy deposited in the calorimeters has to be smeared
 according to an ``additional'' $\Delta E/E$ function, which is different
 for different experimental conditions.
 For example, to match the experimental energy resolution
 for electromagnetic showers in the CB at a low intensity of the photon beam
 (i.e., with few pileups),
 the energy of clusters in the analysis of the MC simulation 
 must be smeared according to $\Delta E/E = 0.0145/(E[\mathrm{GeV}])^{0.34}$.
 Matching the energy resolution between the experimental and
 MC events can be adjusted via reaching
 agreement in the invariant-mass resolution
 and in the kinematic-fit stretch functions (or pulls)
 for the energy of the photons detected in the CB and TAPS.
 The superposition of the inherent
 and additional resolution functions is used in the kinematic-fit
 analysis of both the experimental data and MC simulation. This resolution
 function provides the uncertainties in the energy of the photon clusters.    
 The functions parameterizing the energy resolution
 for electromagnetic showers in the CB and TAPS
 calorimeter are given in the section describing the experimental setup.

 For seven-cluster events, the information from the recoil-proton cluster
 is also used in the kinematic fit. However, in contrast to the treatment
 of a photon cluster, the energy of the
 proton cluster is not included in the fit, as there is a large
 uncertainty in the loss of kinetic energy of the protons
 when they travel from the target to the calorimeter crystals.
 This loss occurs in the material located between the target and
 the crystal surface. The calculation of the kinetic energy of the protons
 from the cluster energies is complicated, as
 the amount of the material depends on the production angles
 in the laboratory system, and the energy loss in matter
 depends on the kinetic energy of the protons. Moreover, when
 the kinetic energy is above 450~MeV for the CB and above 370~MeV for TAPS,
 the recoil protons do not stop inside the crystals. 
 In the analysis of six-cluster events, all parameters of the recoil proton
 (i.e., the kinetic energy and the two angles of the momentum vector)
 are free variables of the kinematic fit.
 The kinematic fitting includes the four main constraints,
 which are based on the conservation of energy and three-momentum,
 and additional constraints on the invariant masses of certain
 particles in the final state. For our reaction, we can use
 three constraints on the invariant mass of two photons
 to have the $\pi^0$-meson mass and a constraint on
 the invariant mass of six photons to have the $\eta$-meson mass.
 Then the total number of constraints to test the hypothesis
 of reaction~(\ref{eqn:eta3pi0}) is eight.
 The effective number of constraints is smaller by the number
 of free variables of the fit, which are the $Z$ coordinate
 of the vertex and the unknown proton parameters (one for seven-cluster
 events and three for six-cluster events).
 Thus the test of hypothesis~(\ref{eqn:eta3pi0}) is a 4-C
 (four effective constraints) fit
 for six-cluster events and a 6-C fit for seven-cluster events.

 Note that testing the $\gamma p \to \eta p \to 3\pi^0 p \to 6\gamma p$
 hypothesis means testing this for all possible permutations of
 pairing the six photon clusters to form three $\pi^0$s.
 For six-cluster events, there are 15 such permutations.   
 In the case of seven-cluster events, this number is seven times larger,
 as the proton cluster is also involved in the permutations.
 The number of permutations can be decreased by the separation
 of the photon clusters from the proton one. In our analysis,
 we used a limit on the $\theta$ angle of clusters, which can be
 only forward ones for the outgoing proton, and the information
 on the time of flight between a TAPS cluster and the CB signal
 with respect to the energy of the TAPS cluster.
 The time-of-flight information was also used to remove a small
 background from the six-cluster events, in which one of the clusters
 was due to the proton. 
 The events for which at least one pairing combination satisfied
 the hypothesis of reaction~(\ref{eqn:eta3pi0}) at the 2\%
 confidence level, CL, (i.e., with a probability greater
 than 2\%) were accepted as $\eta\to 3\pi^0$ event candidates.
 The pairing combination with the largest CL was used
 to reconstruct the kinematics of the reaction.
 A tighter cut on the kinematic-fit CL is unnecessary as
 there is almost no physical background to suppress.
 A looser cut on the CL is not desirable because of inclusion of
 events with poor resolution.

 For the experimental events, in which there are usually
 several tagger hits recorded for one DAQ trigger,
 the $\gamma p \to \eta p \to 3\pi^0 p \to 6\gamma p$ hypothesis
 is tested for every incident photon for which the tagging time is
 within a chosen interval
 and with the incident-photon energy
 above the reaction threshold of 707~MeV.
 If an $\eta\to 3\pi^0$ event candidate from
 one trigger passes the 2\%~CL criterion for several
 tagger hits, they are analyzed as separate events.
 The tagging-time spectrum for the experimental events
 selected as $\eta\to 3\pi^0$ candidates is shown in
 Fig.~\ref{fig:beameta_mamic}(a). The prompt peak
 from the coincidence of the DAQ triggers
 and the tagged photons sits above a uniform
 random background. All the experimental spectra
 were produced from the events where the tagging time
 lay between the vertical lines shown in
 Fig.~\ref{fig:beameta_mamic}(a), but are corrected
 for the random background by subtraction of the spectra
 produced from the events where the tagging time lay
 outside the prompt region.
 To decrease the statistical uncertainties
 and fluctuations in the experimental spectra obtained
 after the random-background subtraction,
 the width of the random region was taken to be much
 wider than the prompt one.
 The normalization factor for the subtraction of
 the random-background spectra is then
 the ratio of the widths taken for the prompt and the random parts
 of the spectrum.

 Another source of background comes from interactions of the incident photons
 with the target walls. This background was investigated by
 analyzing the data taken with an empty target.
 The size of the so-called ``empty-target'' background depends
 on the thickness of the target walls and on how well
 the $\eta\to 3\pi^0$ production on hydrogen can be separated
 from production on heavier nuclei.
 Since the kinematic fit tests the hypothesis of 
 the photon-proton interaction, this rejects many $\eta\to 3\pi^0$
 events produced in the target walls.
 From our analysis, the size
 of the empty-target background in our $\eta\to 3\pi^0$ events
 is 2.2\% after applying a cut at the 2\%~CL of the kinematic fit.  
 Tightening this cut to the 10\%~CL decreases this background to
 1.7\%. Another way to suppress the empty-target background is
 to require the detection of the final-state proton.
 In this case, the empty-target background in our $\eta\to 3\pi^0$ events
 is 0.8\% after applying a cut on the 2\%~CL and 0.65\% for the 10\%~CL
 The corresponding numbers for our experimental $\eta\to 3\pi^0$ events
 with the undetected proton are 7.4\% for the 2\%~CL and
 6.0\% for the 10\%~CL
 Because of limited statistics of the empty-target data,
 the remaining empty-target background was not subtracted from 
 our experimental $\eta\to 3\pi^0$ spectra.
 The results obtained for the slope parameter $\alpha$ 
 by the change in the size of the remaining
 empty-target background showed no dependence on it.
\begin{figure*}
\includegraphics[width=12.cm,height=11.cm,bbllx=.5cm,bblly=.5cm,bburx=19.cm,bbury=16.5cm]{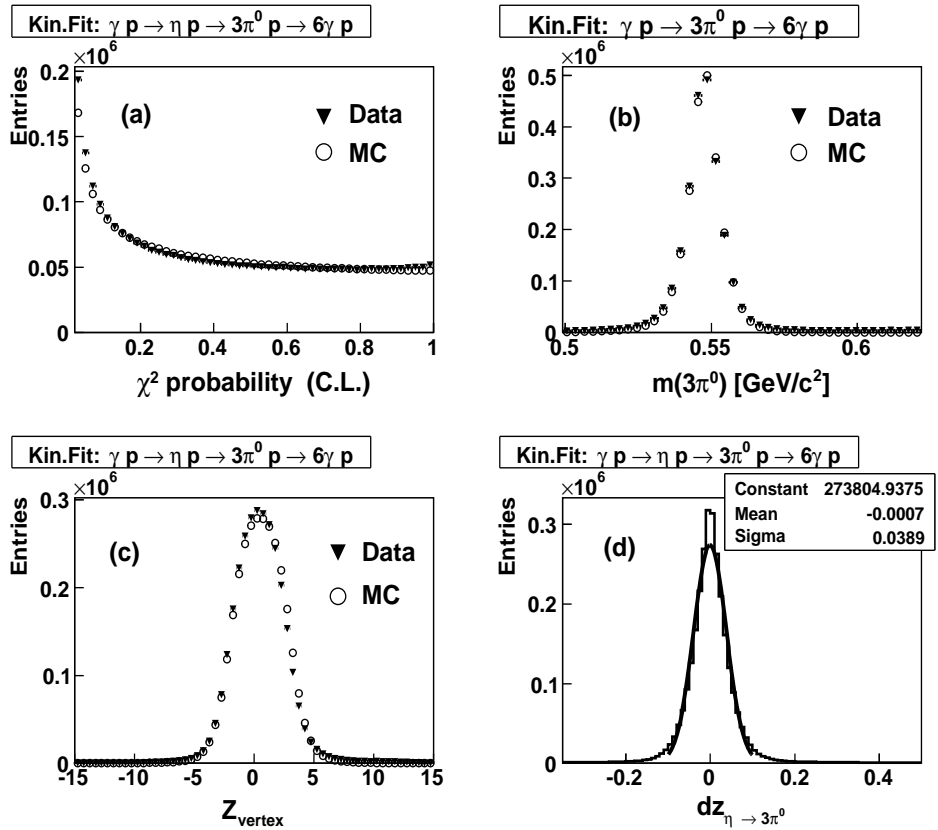}
\caption{
 (a) The $\chi^2$ probability (or CL) distribution for the experimental
 (triangles) and MC (circles) $\eta\to 3\pi^0$ events;
 (b) the $3\pi^0$ invariant mass for the experimental
 (triangles) and MC (circles) $\eta\to 3\pi^0$ events
  selected at the 2\% CL by testing
  the $\gamma p \to 3\pi^0 p \to 6\gamma p$ hypothesis;
 (c) the $Z$ coordinate of the vertex for the experimental
    (triangles) and MC (circles) $\eta\to 3\pi^0$ events;
 (d) the resolution in the variable $z$.
}
 \label{fig:eta3pi0_mamic} 
\end{figure*}

 The MC simulation of the $\gamma p \to \eta p \to 3\pi^0 p$ reaction
 was divided into two steps. First, this reaction was simulated without
 dependence of its yield on the incident-photon energy.
 This MC simulation was
 used to determine the $\gamma p \to \eta p$ excitation function.
 Since we analyze the $\gamma p \to \eta p \to 3\pi^0 p$ reaction
 at a large range of incident-photon energies,
 where the production angular distribution
 changes much, an isotropic distribution was used
 as an input for our MC simulation. The simulation of the $\eta \to 3\pi^0$ decay
 was made according to phase space (i.e., with the slope parameter $\alpha=0$).
 All MC events were propagated through a full {\sc GEANT} (version 3.21)
 simulation of the CB-TAPS detector, folded with resolutions of the detectors
 (such as smearing according to the ``additional'' $\Delta E/E$ function) and
 conditions of the trigger, and analyzed in
 the same way as the experimental data.
 The resulting detector acceptance for
 the $\gamma p \to \eta p \to 3\pi^0 p$ events
 selected by the kinematic fit at the 2\%~CL
 is shown in Fig.~\ref{fig:beameta_mamic}(b); it varies from about
 45\% at the $\eta$ threshold to about 25\% at an incident-photon energy of 1.4~GeV.
 The experimental yield of the $\gamma p \to \eta p \to 3\pi^0 p$ events
 corrected for the acceptance, for the $\eta \to 3\pi^0$
 branching ratio, for the photon beam flux, and for the number of the target protons
 is compared in Fig.~\ref{fig:beameta_mamic}(c)
 to the current PWA-fit solution taken from the SAID~\cite{SAID} data base.
 The shape of the $\gamma p \to \eta p$ excitation function
 at the production threshold confirms the good quality of the
 energy calibration of the tagger, which was performed as explained
 in Ref.~\cite{Tagger2}.
 The small disagreement that is seen at higher energies
 can be partially due to
 the difference between the real production angular distribution and
 the isotropic distribution used in our MC simulation.
 In the second step, the $\gamma p \to \eta p \to 3\pi^0 p$ reaction
 was simulated according to its excitation function folded with
 the bremsstrahlung photon distribution.
 This MC simulation was then used to determine the slope parameter $\alpha$.

 Since the event selection in our analysis is based on the confidence level
 of the kinematic fit,
 it requires good agreement of the $\chi^2$ probability
 (same as CL) distribution for the experimental and MC events.
 Then a change in the cut on
 the CL value removes the same fraction of events from
 the experimental data and MC simulation.
 In Fig.~\ref{fig:eta3pi0_mamic}(a), we compare the CL distributions
 for the experimental (triangles) and MC (circles) events selected by testing
 the $\gamma p \to \eta p \to 3\pi^0 p \to 6\gamma p$ hypothesis.
 These distributions are in reasonable agreement. 
 Note that some increase in the CL distributions for low probability
 is due to events with partially overlapping photon showers, or
 with some leakage of the energy of the showers from the 
 edge crystals of the CB and TAPS.
 Since the energy-resolution functions for the CB and TAPS were determined for
 ``solitary'' electromagnetic showers, for which all energy is deposited
 in the calorimeter, the errors in the energy
 and angles are underestimated for these ``nonideal'' clusters.

 The agreement between the experimental data and MC simulation
 for the energy calibration of the calorimeters
 and the invariant-mass resolution
 can be illustrated by a comparison of the $3\pi^0$ invariant-mass spectra.
 These spectra obtained from events selected by testing
 the $\gamma p \to 3\pi^0 p \to 6\gamma p$ hypothesis are shown 
 in Fig.~\ref{fig:eta3pi0_mamic}(b) by triangles for the experimental data
 and by circles for the MC simulation.
 There is good agreement between the experimental and MC spectrum
 for the mean value, which is consistent with the $\eta$-meson mass of 547.5~MeV,
 and for the invariant-mass resolution, which has $\sigma \approx 6$~MeV.
\begin{figure*}
\includegraphics[width=14.cm,height=5.5cm,bbllx=1.cm,bblly=0.5cm,bburx=19.cm,bbury=7.5cm]{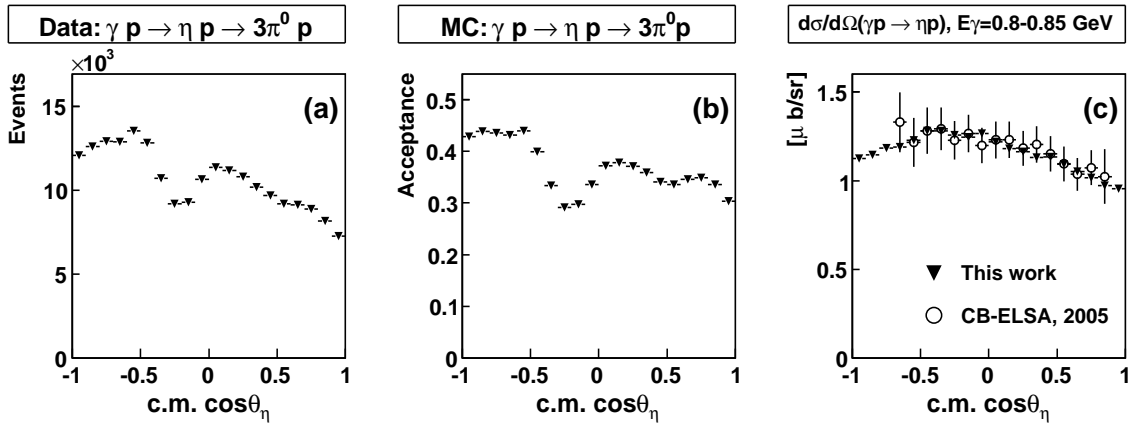}
\caption{
 The angular distribution in the center-of-mass (c.m.) system
 for $\eta$ photoproduction in the $\gamma p \to \eta p$ reaction
 for incident-photon energies of 800 to 850~MeV:
 (a) measured distribution for the $\gamma p \to \eta p \to 3\pi^0 p$ events,
    not yet corrected for the detector acceptance;
 (b) detector acceptance for the $\gamma p \to \eta p \to 3\pi^0 p$ events;
 (c) our $\gamma p \to \eta p$ differential cross sections (triangles)
 compared to the CB-ELSA data~\protect\cite{CB-ELSA} (circles).
}
 \label{fig:costh_eta3pi0_bm3} 
\end{figure*}

 Since the $Z$ coordinate of the vertex is a free variable
 in the kinematic fit, the $Z$ distribution must reflect the thickness of
 the LH$_2$ target, which is 5 cm long, and the target position.
 The agreement of these distributions for the experimental (triangles) and
 MC (circles) $\eta\to 3\pi^0$ events can be seen 
 in Fig.~\ref{fig:eta3pi0_mamic}(c).
 The larger width of the $Z$ distribution, compared to 
 the 5-cm thickness of the physical target,
 is due to the resolution of the kinematic fit in the $Z$ coordinate.
 For the $\gamma p \to \eta p \to 3\pi^0 p \to 6\gamma p$ process, this
 resolution is about 1.1~cm; it is determined from the difference
 between the initial and reconstructed value of $Z$ in the MC simulation.   
\begin{figure*}
\includegraphics[width=14.cm,height=5.5cm,bbllx=1.cm,bblly=0.5cm,bburx=19.cm,bbury=7.5cm]{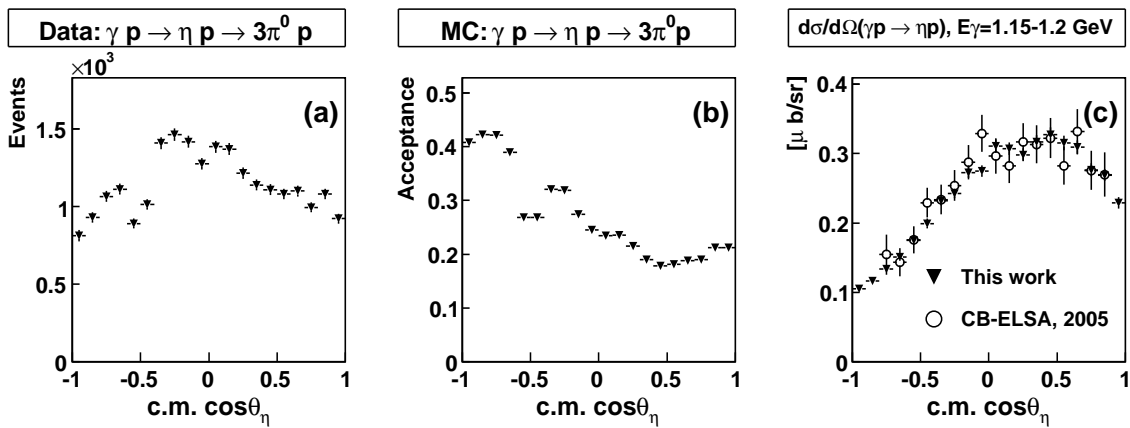}
\caption{
 Same as Fig.~\protect\ref{fig:costh_eta3pi0_bm3} but for
 incident-photon energies of 1150 to 1200~MeV.
}
 \label{fig:costh_eta3pi0_bm10} 
\end{figure*}

 The resolution in the variable $z$, defined in Eq.(\ref{eqn:zvar}),
 can be understood from a comparison of $z$ calculated from
 the kinematics of the initially simulated events and $z$ calculated from
 the kinematics reconstructed for these events by the kinematic fit.
 The resolution in $z$ is shown in Fig.~\ref{fig:eta3pi0_mamic}(d).
 Based on this resolution and the $z$-variable limits, which are from 0 to 1,
 we divided our $z$ spectra, used for the determination of the slope parameter,
 into 20 bins. This binning provides a bin width that
 is wider than one $\sigma_z=0.039$.  
 The spectrum shown in Fig.~\ref{fig:eta3pi0_mamic}(d) also demonstrates
 the insignificance of the combinatorial background in
 $\eta\to 3\pi^0\to 6\gamma$ events
 (i.e., when the kinematic-fit hypothesis with the largest CL
 chooses a false pairing combination of the six photons to the three $\pi^0$s).
 The combinatorial background produces accidental reconstructed
 values for $z$ that are spread widely in the $dz$ spectrum, resulting
 in a washout of the real Dalitz-plot slope.    
 For a rough estimate of the combinatorial-background contribution,
 we took the fraction of the MC events that have $|dz|>0.2$;
 this fraction was found to be 3.5\% only.  
 
 To show that our MC simulation satisfactorily reproduces the experimental
 acceptance, we compare our differential cross sections for
 the $\gamma p \to \eta p$ reaction with other
 existing measurements. In Fig.~\ref{fig:costh_eta3pi0_bm3}(a), we show
 the experimental distribution of the $\eta$ production angle
 from $\gamma p \to \eta p \to 3\pi^0 p$ events measured
 in the center-of-mass (c.m.) system for incident-photon
 energies of 800 to 850~MeV.
 This distribution is not yet corrected for the acceptance.
 The corresponding acceptance obtained from our MC simulation
 is shown in Fig.~\ref{fig:costh_eta3pi0_bm3}(b). 
 The complicated behavior of both the experimental distribution
 and the acceptance is mostly caused by the gap between the CB and TAPS calorimeters.
 The experimental distribution corrected for the acceptance, for the $\eta \to 3\pi^0$
 branching ratio, for the photon beam flux, and for the number of the target protons
 is shown by triangles in Fig.~\ref{fig:costh_eta3pi0_bm3}(c). This distribution
 has a smooth shape now, which is in good agreement with
 the $\gamma p \to \eta p$ differential cross section obtained recently
 for the same energy range by the Crystal Barrel Collaboration
 at ELSA (CB-ELSA)~\cite{CB-ELSA}. The CB-ELSA data are shown by circles
 in the same figure.
 Similarly good agreement with the corresponding CB-ELSA results is observed
 for all other energy intervals. In Fig.~\ref{fig:costh_eta3pi0_bm10},
 we illustrate it just for one more energy interval of 1150 to 1200~MeV.
 The agreement in the differential
 cross sections is shown to demonstrate the quality of our $\eta \to 3\pi^0$ analysis. 

\section{Determination of the slope parameter $\alpha$ and its uncertainty}

\begin{figure*}
\includegraphics[width=13.cm,height=10.5cm,bbllx=0.5cm,bblly=0.5cm,bburx=19.cm,bbury=17.cm]{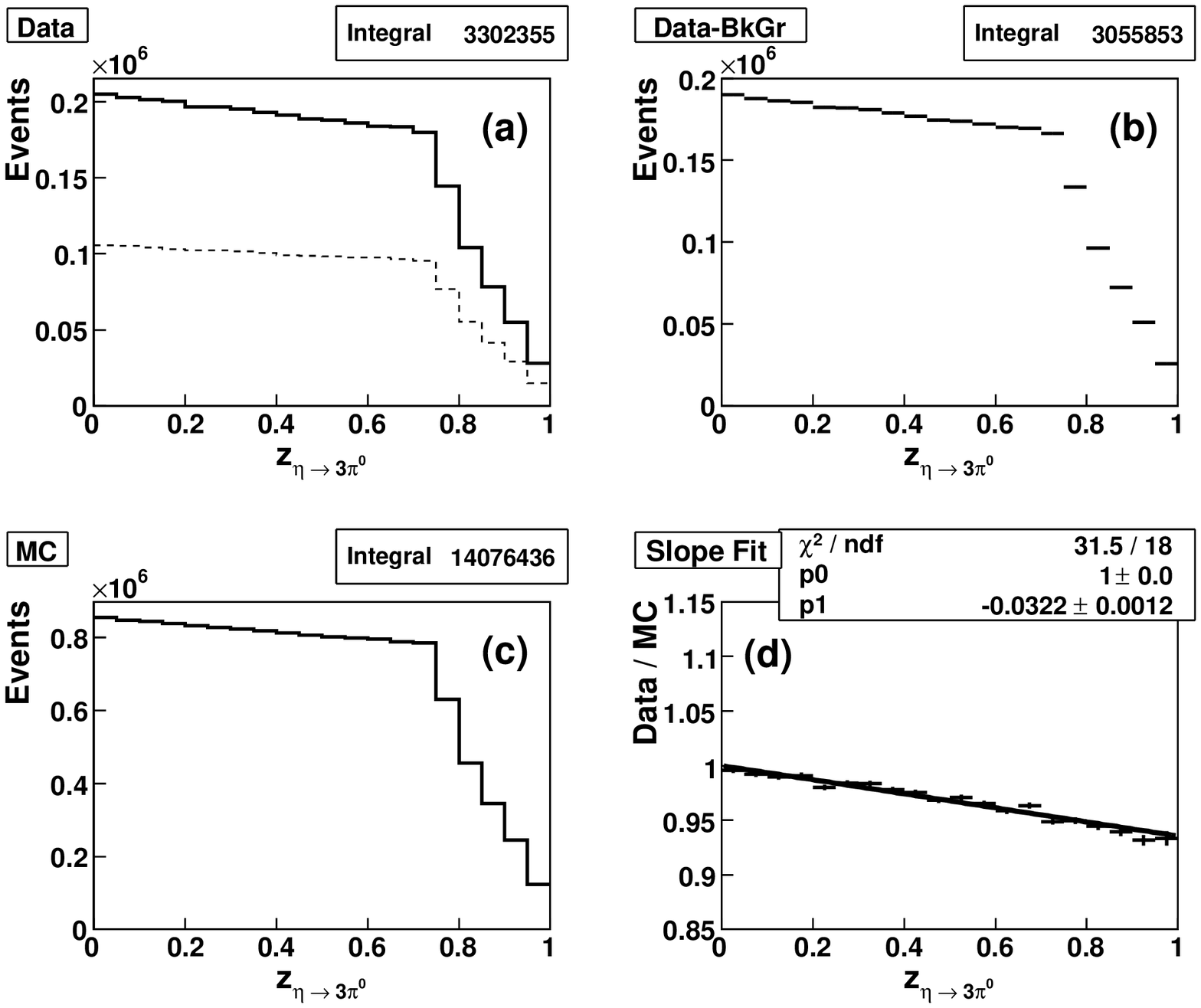}
\caption{
 The $z$ distribution for the $\eta\to 3\pi^0$ events selected at CL=2\%
 by testing the $\gamma p \to \eta p \to 3\pi^0 p \to 6\gamma p$ hypothesis:
 (a) experimental prompt (solid line) and unnormalized random (dashed line) data;
 (b) experimental data after the subtraction of the normalized random background;
 (c) MC simulation for $4\times10^7$ $\gamma p \to \eta p \to 3\pi^0 p$ events;
 (d) the ratio of the experimental and MC distributions fitted
     to the function $p_0 + 2 p_1 z$. 
}
 \label{fig:slope_mamic_all} 
\end{figure*}

 The experimental $z$ distribution obtained for the $\eta\to 3\pi^0$ events
 with prompt tagging times is shown as the solid line
 in Fig.~\ref{fig:slope_mamic_all}(a).
 The events were selected at the 2\%~CL from the test of
 the $\gamma p \to \eta p \to 3\pi^0 p \to 6\gamma p$ hypothesis with the kinematic fit.
 Both six- and seven-cluster events (i.e., without and with the outgoing proton detected)
 were included in this distribution; the fraction of
 six-cluster events is only about 20\%.
 The corresponding (unnormalized) $z$ distribution obtained from the events
 with the random tagging times is shown
 in the same figure by the dashed line.
 The size of this background depends on the photon-beam intensity
 and is less than 10\% for our data. 
 The experimental $z$ distribution after subtraction of the normalized
 random background is shown in Fig.~\ref{fig:slope_mamic_all}(b).
 The $z$ distribution obtained from our MC simulation
 is shown in Fig.~\ref{fig:slope_mamic_all}(c).
 This MC simulation is based on $4\times10^7$ $\gamma p \to \eta p \to 3\pi^0 p$
 events with $\eta$ decaying to $3\pi^0$ according to phase space (i.e., with $\alpha=0$).
 The ratio of the experimental $z$ distribution
 to the MC one that was fitted to the function $p_0 + 2 p_1 z$ is
 shown in Fig.~\ref{fig:slope_mamic_all}(d). 
 To bring this fit to the required function $1 + 2\alpha z$,
 the MC distribution was normalized in such a way as to make the fit parameter
 $p_0$ equal to 1. Then the fit parameter $p_1=-0.0322\pm0.0012$ has the meaning of
 the slope parameter $\alpha$. We take the error of $p_1$ from this fit, in which our full
 set of experimental statistics was used, as the statistical uncertainty of 
 the slope parameter $\alpha$. To estimate its systematic uncertainty,
 a variety of tests were performed; these are listed in Table~\ref{tab:rslt1}.

 For each test listed in Table~\ref{tab:rslt1}, we include
 information on the criteria for event selection,
 the experimental statistics, the value for the slope parameter
 with its statistical uncertainty,
 and the $\chi^2$/ndf value of the fit. The result of the fit shown
 in Fig.~\ref{fig:slope_mamic_all}(d) is listed in Table~\ref{tab:rslt1} as test \#1.
 Tests \#2---\#4 check the sensitivity of the results to the cut on the CL of the
 kinematic fit. Tightening the cut on the CL results in a data set with better
 resolution and less remaining background. The variation in the value for $\alpha$
 is much smaller than the statistical uncertainties.

\begin{figure*}
\includegraphics[width=13.cm,height=5.5cm,bbllx=1.cm,bblly=0.5cm,bburx=19.cm,bbury=8.5cm]{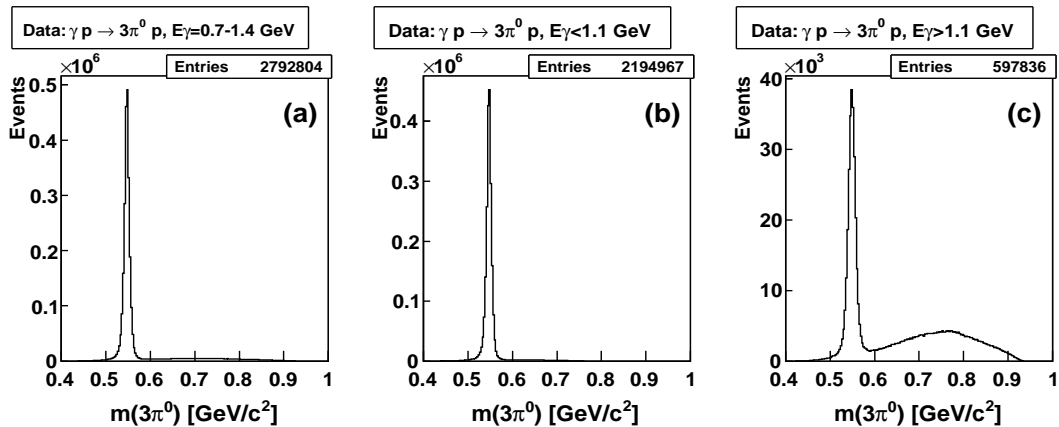}
\caption{
 Experimental spectra for the $3\pi^0$ invariant mass obtained by testing
 the $\gamma p \to 3\pi^0 p \to 6\gamma p$ hypothesis at the 2\% CL for
 (a) the full tagging range of the incident-photon energies,
 $E_g=0.7-1.4$~GeV, (b) $E_g=0.7-1.1$~GeV, and (c) $E_g=1.1-1.4$~GeV. 
}
 \label{fig:m3pi0_bm} 
\end{figure*}

 Tests \#5---\#7 check the sensitivity of the results to a possible background
 from the $3\pi^0$ final state that is not due to $\eta$ decay.
 The size of this background can be understood from the examination of
 the $3\pi^0$ invariant-mass spectrum obtained when testing
 the $\gamma p \to 3\pi^0 p$ hypothesis.
 This spectrum is shown in Fig.~\ref{fig:m3pi0_bm}
 for the full tagging range 0.7---1.4~GeV of incident-photon energies
 and for the ranges 0.7---1.1~GeV and 1.1---1.4~GeV.
 An examination of the spectrum shows that the size of
 the $3\pi^0$ background comprises a few percent
 of our full data set, and it is negligibly small for events with
 the incident-photon energies below 1.1~GeV.
 We tested three cuts on the incident-photon energy: 0.9, 1.0, and 1.1~GeV.
 The experimental value for $\alpha$ is almost independent of this cut.

 Tests \#8 and \#9 check whether the simulation
 of the threshold on the CB total energy in the DAQ trigger
 is correct. From the analysis of the sum of the cluster energies,
 the parameters of the trigger were determined to be 320~MeV
 for the threshold itself and $\sigma=20$~MeV for its uncertainty.
 To reproduce these trigger conditions in the MC analysis,
 the total energy of clusters has be smeared according to a normal
 distribution having that $\sigma$, and those events that have
 a smeared energy less than the threshold value must be rejected from
 the analysis.
 In our tests, we applied software thresholds of 420
 and 470~MeV to both the experimental and MC events.
 These magnitudes were chosen to be considerably larger
 than the threshold of 320~MeV smeared with $\sigma=20$~MeV.
 The results obtained for $\alpha$ are in good agreement
 within their statistical uncertainties.
 
  Tests \#10 and \#11 check the sensitivity of our results to the difference between
 the isotropic production angular distribution used in the MC simulation of   
 the $\gamma p \to \eta p$ reaction and the real distributions defined
 by the differential cross sections, which
 depend on the incident-photon energy.
 Examples of the $\gamma p \to \eta p$ differential cross sections
 for the two different intervals of the incident-photon energy
 are shown in Figs.~\ref{fig:costh_eta3pi0_bm3} and \ref{fig:costh_eta3pi0_bm10}.
 In these tests,
 we determined the slope parameter from two subsets. The first subset included
 the events with only $\cos \theta_{\eta}>0$
 and the second one with only $\cos \theta_{\eta}<0$.
 Both the results for $\alpha$ are in agreement
 within their statistical uncertainties.

 In tests \#12---\#17, we repeated some of the previous tests but for
 the seven-cluster events only. The results obtained are in good agreement
 with each other and with the tests that were performed
 using the sum of the events with both the cluster multiplicities.

 Tests \#18---\#20 are performed for the six-cluster events only.
 The results for $\alpha$ are slightly smaller here in comparison
 to the corresponding seven-cluster results. This could be in part due to
 a larger fraction of the empty-target background in the six-cluster events.
 The $\eta\to 3\pi^0$ decay kinematics for this background is
 smeared by the kinematic fit, which assumes the target to be a proton.
 This smearing leads to poorer resolution in the variable $z$ and
 increases the combinatorial background, which reduces
 the real slope of the $\eta\to 3\pi^0$ Dalitz plot.
 A similar smearing occurs for the $\eta\to 3\pi^0$ events with
 accidental incident-photon energies (i.e., our random-background
 events). Test \#21, in which $\alpha$ is obtained from
 the $\eta\to 3\pi^0$ events with accidental incident photons,
 illustrates the reduction in the experimental slope
 caused by the smearing effect.
 Differently from the empty-target background, the
 random background was subtracted in all tests of Table~\ref{tab:rslt1},
 except test \#21.

 Tests \#22---\#24 illustrate the stability of the results over the period
 of data taking. Our data set includes three periods of data taking from April 2007
 to July 2007 with similar experimental and trigger conditions but different durations.
 All three results are in good agreement within their statistical uncertainties.
\begin{table*}
\caption
[tab:rslt1]{
 Results for the $\eta\to 3\pi^0$ slope parameter $\alpha$ 
 for different selection criteria
 } \label{tab:rslt1}
\begin{ruledtabular}
\begin{tabular}{|c|l|c|c|c|} 
 Result \# & Cuts & Statistics & $\alpha$ & $\chi^2$/ndf \\
\hline
 1   & CL=2\% & $3.06\times10^6$ & $-0.0322\pm 0.0012$ & 31.5/18 \\
 2   & CL=5\% & $2.78\times10^6$ & $-0.0326\pm 0.0013$ & 32.2/18 \\
 3   & CL=10\% & $2.50\times10^6$ & $-0.0329\pm 0.0014$ & 30.0/18 \\
 4   & CL=20\% & $2.11\times10^6$ & $-0.0326\pm 0.0015$ & 25.9/18 \\
 5   & CL=2\%, $E_{\gamma}< 1.1$~GeV & $2.76\times10^6$ & $-0.0320\pm 0.0013$ & 26.9/18 \\
 6   & CL=2\%, $E_{\gamma}< 1.0$~GeV & $2.58\times10^6$ & $-0.0320\pm 0.0013$ & 28.9/18 \\
 7   & CL=2\%, $E_{\gamma}< 0.9$~GeV & $2.18\times10^6$ & $-0.0321\pm 0.0015$ & 20.2/18 \\
 8   & CL=2\%, $E_{\mathrm{CB}}< 0.42$~GeV & $2.83\times10^6$ & $-0.0316\pm 0.0013$ & 29.1/18 \\
 9   & CL=2\%, $E_{\mathrm{CB}}< 0.47$~GeV & $2.60\times10^6$ & $-0.0319\pm 0.0013$ & 30.7/18 \\
 10   & CL=2\%, c.m. $\cos\theta_{\eta}< 0.$ & $1.73\times10^6$ & $-0.0334\pm 0.0017$ & 23.7/18 \\
 11   & CL=2\%, c.m. $\cos\theta_{\eta}> 0.$ & $1.32\times10^6$ & $-0.0312\pm 0.0019$ & 14.5/18 \\
 12  & CL=2\%, 7 cl. & $2.39\times10^6$ & $-0.0323\pm 0.0014$ & 26.4/18 \\
 13  & CL=10\%, 7 cl. & $1.97\times10^6$ & $-0.0327\pm 0.0015$ & 27.8/18 \\
 14  & CL=20\%, 7 cl. & $1.67\times10^6$ & $-0.0325\pm 0.0016$ & 26.9/18 \\
 15  & CL=2\%, 7 cl., $E_{\gamma}< 1.1$~GeV& $2.13\times10^6$ & $-0.0319\pm 0.0015$ & 25.9/18 \\
 16  & CL=2\%, 7 cl., $E_{\gamma}< 1.0$~GeV& $1.97\times10^6$ & $-0.0319\pm 0.0015$ & 28.5/18 \\
 17  & CL=2\%, 7 cl., $E_{\gamma}< 0.9$~GeV& $1.62\times10^6$ & $-0.0323\pm 0.0017$ & 23.4/18 \\
 18  & CL=2\%, 6 cl. & $0.663\times10^6$ & $-0.0292\pm 0.0027$ & 22.0/18 \\
 19  & CL=10\%, 6 cl. & $0.525\times10^6$ & $-0.0307\pm 0.0030$ & 29.3/18 \\
 20  & CL=20\%, 6 cl. & $0.433\times10^6$ & $-0.0301\pm 0.0033$ & 25.4/18 \\
 21  & CL=2\%, random BkGr & $1.73\times10^6$ & $-0.0132\pm 0.0016$ & 23.2/18 \\
 22   & CL=2\%, ~04.2007 & $0.617\times10^6$ & $-0.0308\pm 0.0026$ & 21.0/18 \\
 23   & CL=2\%, ~06.2007 & $1.50\times10^6$ & $-0.0324\pm 0.0017$ & 23.6/18 \\
 24   & CL=2\%, ~07.2007 & $0.939\times10^6$ & $-0.0329\pm 0.0021$ & 16.8/18 \\
\end{tabular}
\end{ruledtabular}
\end{table*}

 The uncertainty in measuring the parameter $\alpha$ due to our
 limited resolution in the variable $z$ and the combinatorial
 background can be estimated by
 introducing the measured $\alpha$ value into the MC simulation
 and using this MC sample instead of the experimental data.
 To exclude
 statistical fluctuations from this estimate, the same MC sample
 must be used in the ratio of the simulations with nonzero and zero
 $\alpha$. This can be done when an event enters into the $z$ spectrum
 according to the $z$ value that is reconstructed by our program but 
 with a weight of $1+2\alpha z$, calculated by using $z$ from the simulated
 kinematics of the event. In this way, the ideal $z$ distributions are folded
 with the experimental acceptance and resolution, and
 both spectra that are used in the ratio
 have correlated statistical fluctuations.
 These fluctuations are canceled in the ratio and do not smear
 the magnitude of the slope.
 The use of $\alpha=-0.032$ as an input for this estimate resulted in
 $\alpha=-0.0300\pm 0.0007$ after applying the criteria of test~\#1 and
 $\alpha=-0.0303\pm 0.0009$ after applying the criteria of test~\#4.
 The difference 0.0017 between the input $\alpha$
 and the value obtained for $\alpha$ using the criteria of test~\#4
 is slightly smaller than the value obtained using the criteria of test~\#1.
 This is expected because of a tighter quality cut
 on the CL of the kinematic fit in test~\#4.
 In the analysis of the MC simulation, we can also decrease
 the combinatorial background by applying an additional cut on
 the difference between the initially simulated
 and reconstructed value of $z$.
 The elimination of the MC events that have $|dz|>0.2$,
 additionally to the selection criteria of test~\#1,
 results in $\alpha=-0.0315\pm 0.0007$, which is
 in good agreement with $\alpha=-0.032$ used as an input.
 These tests illustrate the importance of the experimental
 resolution for the precision measurement of
 the slope parameter $\alpha$.

 As our final result for the slope parameter $\alpha$,
 we use the weighted average of all results from Table~\ref{tab:rslt1}
 except tests \#18---\#21, which were performed for
 the six-cluster events and the random background.
 The six-cluster results are omitted as they involve larger background
 and much poorer statistics.
 The weight factor of each result was taken as the inverse value
 of its statistical uncertainty.
 This procedure gives $-0.0322$ for the value of $\alpha$.
 Note that this value is identical to the result for $\alpha$
 obtained from test \#1, which is based on our full experimental
 statistics and could be considered as an alternative choice
 of our main value for $\alpha$.
 For the statistical uncertainty of $\alpha$,
 we take the uncertainty 0.0012 from our full-statistics test \#1.
 In the systematic uncertainty of $\alpha$, we include half of the largest variation
 of the results in Table~\ref{tab:rslt1} (except tests \#18---\#21),
 which is 0.0013, and 0.0017 obtained earlier as the difference between
 the initially simulated $\alpha=-0.032$ and the reconstructed value for $\alpha$.
 Adding these uncertainties in quadrature gives 0.0022 for our
 total systematic uncertainty.

 An uncertainty in the parameter $\alpha$ resulting from a possible cusp in
 the $\pi^0\pi^0$ invariant-mass spectrum is found to be insignificant.
 The details on the cusp search are given in the next section. 

 The value
 $\alpha=-0.0322\pm0.0012_{\mathrm{stat}}\pm0.0022_{\mathrm{syst}}=
 -0.0322\pm0.0025_{\mathrm{tot}}$,
 obtained in our measurement for the $\eta\to 3\pi^0$ slope parameter ,
 is in good agreement with the PDG value for $\alpha$,
 which is $-0.031\pm0.004$~\cite{PDG}, but it has a smaller uncertainty.
 That the present PDG value for $\alpha$ is identical
 to the result of the Crystal Ball at the AGS~\cite{eta_slope_bnl}
 means that the new Crystal Ball Collaboration at
 MAMI confirms the previous measurement. 
 Finally, taking into account the magnitude of our total uncertainty,
 we prefer to give the rounded value
\begin{equation}
 \alpha=-0.032\pm0.003_{\mathrm{tot}}
 \label{eqn:valpha}
\end{equation}
 as our final value for the slope parameter.

\section{Search for a cusp structure in the $\pi^0\pi^0$ invariant-mass spectrum}

\begin{figure*}
\includegraphics[width=13.5cm,height=6.cm,bbllx=1.cm,bblly=0.5cm,bburx=19.cm,bbury=9.cm]{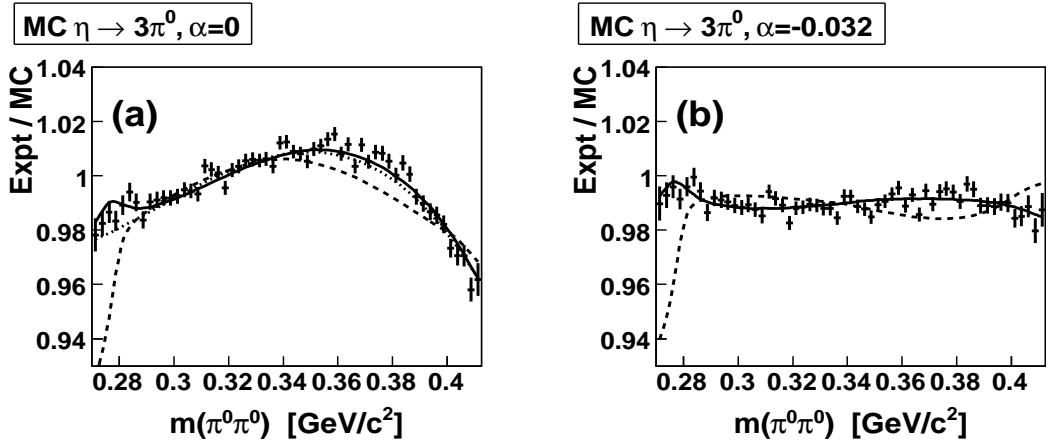}
\caption{
 The ratio of the experimental spectrum
 for the $\pi^0\pi^0$ invariant mass to the corresponding MC spectrum
 is shown by crosses for
 (a) the phase-space decay of $\eta\to 3\pi^0$ in the MC simulation (i.e., with $\alpha=0$)
 and for (b) the MC simulation of the $\eta\to 3\pi^0$ decay with $\alpha=-0.032$~.
  The ratio of the MC simulation with $\alpha=-0.032$ to the MC simulation with $\alpha=0$ is
  shown in (a) by dots. The corresponding ratios of the MC spectra
 with two versions of the cusp prediction are shown by dashed
 (using Ref.~\protect\cite{Belina}) and solid
 (using Ref.~\protect\cite{Bissegger}) lines
 in (a) for the case when $\alpha=-0.032$ and in (b) when $\alpha=0$. 
}
 \label{fig:cusp_eta3pi0} 
\end{figure*}

 The magnitudes of $\chi^2$/ndf for the fits from Table~\ref{tab:rslt1} hint that
 the linear hypothesis, $1+2\alpha z$, is not completely accurate for describing
 the slope of the experimental $\eta\to 3\pi^0$ Dalitz plot.
 A possible explanation could
 be the occurrence of a cusp effect in the $\pi^0\pi^0$ invariant-mass spectrum
 from the $\eta\to 3\pi^0$ decays.
 The origin of such a cusp was
 discussed in the Introduction. In this section,
 we check whether a cusplike structure is seen in our data and
 how its presence could affect our result for $\alpha$.
 Since the effect of the cusp in $\eta\to 3\pi^0$ is expected to be small,
 the most convenient distribution to see it
 is the ratio of the experimental and MC spectrum for the $\pi^0\pi^0$ invariant mass.
 This ratio is shown by crosses in Fig.~\ref{fig:cusp_eta3pi0}(a)
 for the case where the MC simulation of the $\eta\to 3\pi^0$ decay is made
 according to phase space (i.e., with $\alpha=0$).
 Since the experimental $\pi^0\pi^0$ invariant-mass spectrum is already
 distorted from phase space by the nonzero $\alpha$, we also show
 the ratio in which the MC simulation with $\alpha=-0.032$ is taken
 instead of the experimental data; it is shown by dots in the same figure.
 This ratio, for the most part, is in reasonable agreement with the experimental one.
 The largest deviation is seen around a mass of 0.28~GeV, which 
 corresponds to the $\pi^+\pi^-$ threshold.
 To better separate the effect of the cusp from that of a nonzero $\alpha$ value,
 we replaced the phase-space MC simulation in the data-to-MC ratio by the MC
 simulation that 
 includes the slope parameter $\alpha=-0.032$ for the $\eta\to 3\pi^0$ decay.
 The result of this replacement
 is shown by crosses in Fig.~\ref{fig:cusp_eta3pi0}(b).

 The magnitude of the cusplike structure, which is seen at the level of 1\% only,
 is much smaller than the prediction made in Ref.~\cite{Belina}
 (shown by a dashed curve in Fig.~\ref{fig:cusp_eta3pi0}).
 The shape and the sign of the cusplike structure also look different. 
 For a more correct comparison, each
 theoretical prediction shown in the figures
 is folded with our experimental resolution.
 Because of the smallness of the effect observed in the $\pi^0\pi^0$ invariant-mass
 spectrum, better statistics is desirable to draw any final conclusion
 on the magnitude and features of the cusp.

\begin{figure*}
\includegraphics[width=13.5cm,height=6.cm,bbllx=1.cm,bblly=0.5cm,bburx=19.cm,bbury=9.cm]{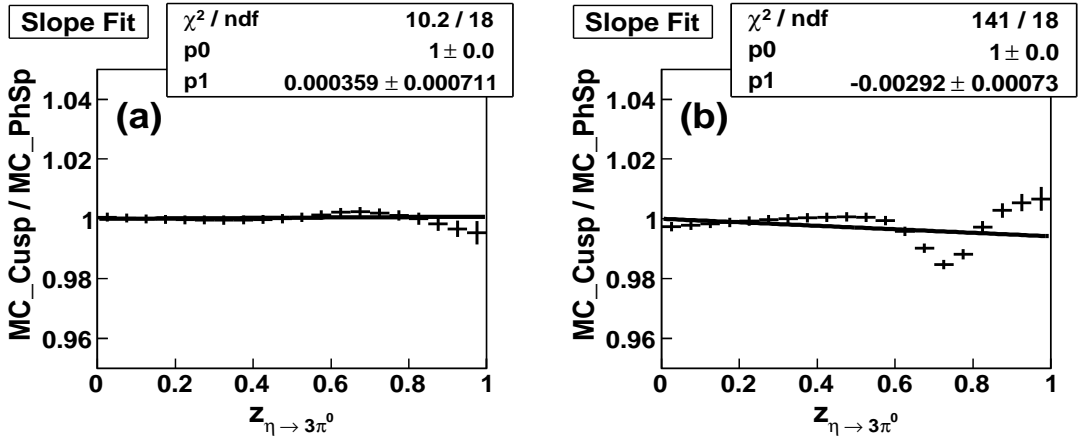}
\caption{
 Ratio of $z$ distributions (crosses) for the $\eta\to 3\pi^0$ decays
 simulated with and without a cusp: (a) corresponds to the
 solid curve and (b) to the dashed curve in Fig.~\protect\ref{fig:cusp_eta3pi0}(b).
 The solid lines show fits to the function $p_0 + 2 p_1 z$.
}
 \label{fig:slope_mc2} 
\end{figure*}
 Trying to reach a closer agreement between the data and a model
 prediction, we used formulas from Ref.~\cite{Bissegger}.
 The authors of this article conduct a detailed theoretical study of cusps
 in $K_L \to 3\pi$ decays and also indicate the way
 to use their approach in the $\eta \to 3\pi$ case if
 one starts from the following parametrization of the
 decay tree amplitudes: $A(\eta \to 3\pi^0) = u_0 + u_1 z$ and
  $A(\eta \to \pi^+\pi^-\pi^0) = v_0 + v_1 y + v_2 y^2 + v_2 x^2$
 with the conventional $\eta \to 3\pi$ Dalitz plot variables.
 The cusp shape in $\eta \to 3\pi^0$ then depends strongly
 on the parameters of the $\eta \to \pi^+\pi^-\pi^0$ tree amplitude,
 which are still not well determined.
 To use the parameters available from analyses of 
 the density distribution of the $\eta\to 3\pi$ Dalitz plots,
 we assumed that 
 $|A(\eta\to 3\pi^0)|^2 \sim 1 + 2(u_1/u_0) z = 1+ 2\alpha z$
 and 
  $|A(\eta \to \pi^+\pi^-\pi^0)|^2 \sim 1 + 2(v_1/v_0) y
  + (2 v_2/v_0 + v_1^2/v_0^2) y^2 + 2(v_3/v_0) x^2 =
  1 + a y + b y^2 + c x^2$.
  It turned out that, after some variation
 of the $\eta \to \pi^+\pi^-\pi^0$ parameters ($a$, $b$,
 and $c$), by starting from the ones that are available in the PDG~\cite{PDG}, 
 the agreement of the cusp prediction with the data can be improved.
 An example of such an improvement is shown by the solid curve
 in Fig.~\ref{fig:cusp_eta3pi0}. Further improvement requires
 a simultaneous fit of the $\eta \to 3\pi^0$ and
 $\eta \to \pi^+\pi^-\pi^0$ data, which is not the topic of this
 paper. 
 The current agreement is sufficient to understand our systematic
 uncertainty in the parameter $\alpha$ resulting from a possible cusp
 at the $\pi^+\pi^-$ threshold.
 In Fig.~\ref{fig:slope_mc2}, we illustrate the deviation of
 the slope in the $z$ distribution from the phase space
 for the two cusp shapes
 that were shown in Fig.~\ref{fig:cusp_eta3pi0}.
 As shown, if the cusp structure is similar to the one that is consistent
 with our data, then the effect of the cusp on
 the $\eta \to 3\pi^0$ slope is negligible.
 However, the cusp as predicted in Ref.~\cite{Belina}
 would cause significant distortion away from the linear behavior;
 this is not observed in our experimental data. 

\section{Summary and conclusions}
 The dynamics of the $\eta\to 3\pi^0$ decay have been
 studied with the Crystal Ball multiphoton spectrometer
 and the TAPS calorimeter. Bremsstrahlung photons produced
 by the 1.5-GeV electron beam of the Mainz Microtron MAMI-C
 and tagged by the Glasgow photon spectrometer were used
 for $\eta$-meson production.
 The analysis of $3\times 10^6$
 $\gamma p \to \eta p \to 3\pi^0 p \to 6\gamma p$ events
 yields the value $\alpha=-0.032\pm0.003$ for the $\eta\to 3\pi^0$
 slope parameter, which agrees with the majority of recent
 experimental results and has the smallest uncertainty.
 The agreement with $\alpha=-0.031\pm0.004$ of the measurement
 made by the Crystal Ball at the AGS,
 where the $\pi^- p \to \eta n$ reaction was used,
 demonstrates the suitability of photoproduction
 for studying subtle effects in $\eta$ decays.
 The $\pi^0\pi^0$ invariant-mass spectrum was investigated
 for the occurrence of a cusplike structure
 in the vicinity of the $\pi^+\pi^-$ threshold.
 The observed effect is small and does not affect
 our measured value for the slope parameter.
 Further investigation of the cusp
 requires better statistics and a simultaneous analysis of
 $\eta\to 3\pi^0$ and $\eta \to \pi^+\pi^-\pi^0$ data.

\begin{acknowledgments}
 The authors thank J. Gasser and A. Fuhrer for their help
 with the cusp simulation. The support from
 the accelerator group of MAMI is much appreciated.
 We thank the undergraduate students of Mount Allison
 and George Washington Universities for their assistance.
 This work was supported by U.S. 
 DOE and NSF (Grant No. PHY 0652549),
 EPSRC and STFC of the United Kingdom,
 the Deutsche Forschungsgemeinschaft
 (SFB 443, SFB/TR 16), the European
 Community-Research Infrastructure Activity under
 the FP6 "Structuring the European Research Area"
 programme (Hadron Physics, Contract No. RII3-CT-2004-506078),
 DFG-RFBR (Grant No. 09-02-91330) of Germany and Russia,
 SNF of Switzerland, and NSERC of Canada.
\end{acknowledgments}

\end{document}